# Empirical Temperature- and Extinction-dependent Extinction Coefficients for the GALEX, Pan-STARRS 1, Gaia, SDSS, 2MASS, and WISE Passbands

Ruoyi Zhang[1,2] and Haibo Yuan[1,2]
[1] Institute for Frontiers in Astronomy and Astrophysics, Beijing Normal University, Beijing 102206, People's Republic of China; yuanhb@bnu.edu.cn
[2] Department of Astronomy, Beijing Normal University, No. 19, Xinjiekouwai St., Haidian District, Beijing, 100875, People's Republic of China
Received 2022 July 18; revised 2022 October 24; accepted 2022 October 25; published 2022 December 23

## Abstract

We have obtained accurate dust reddening from the far-ultraviolet to the mid-infrared for up to 5 million stars by the star-pair algorithm based on LAMOST stellar parameters along with Galaxy Evolution Explorer, Pan-STARRS 1, Gaia, Sloan Digital Sky Survey, Two Micron All Sky Survey, and Wide-field Infrared Survey Explorer photometric data. The typical errors are between 0.01 and 0.03 mag for most colors. We derived the empirical reddening coefficients for 21 colors both in the traditional (single-valued) way and as a function of $T_{eff}$ and $E(B - V)$ by using the largest samples of accurate reddening measurements, together with the extinction values from Schlegel et al. The corresponding extinction coefficients have also been obtained. The results are compared with model predictions and generally in good agreement. Comparisons with measurements in the literature show that the $T_{eff}$- and $E(B - V)$-dependent coefficients explain the discrepancies between different measurements naturally, i.e., using sample stars of different temperatures and reddening. Our coefficients are mostly valid in the extinction range of 0–0.5 mag and the temperature range of 4000–10,000 K. We recommend that the new $T_{eff}$- and $E(B - V)$-dependent reddening and extinction coefficients should be used in the future. A Python package is also provided for the usage of the coefficients (https://github.com/vnohhf/extinction_coeffcient/).

*Unified Astronomy Thesaurus concepts:* Interstellar dust (836); Interstellar dust extinction (837)

## 1. Introduction

Dust is the product of stellar winds or supernova explosions with complex components. By absorption and scattering, dust grains cause wavelength-dependent attenuation of starlight (Draine 2003). Dust extinction and reddening correction play a pivotal role in revealing the intrinsic properties of astronomical objects, thus affecting almost every field in modern astronomy.

With the rapid development of data-processing and calibration techniques (e.g., Padmanabhan et al. 2008; Yuan et al. 2015; Burke et al. 2018; Huang & Yuan 2022), many photometric surveys are able to deliver magnitudes and colors to a precision of approximately 1% for numerous targets. Upcoming surveys, such as the Vera C. Rubin Observatory Legacy Survey of Space and Time (LSST; Ivezić et al. 2019) and the Chinese Space Station Telescope (CSST; Zhan 2018), are going to provide even better photometric data. In particular, in the era of Gaia (Gaia Collaboration et al. 2016, 2018, 2021) millimagnitude-precision photometric data are available for numerous stars, as are microarcsecond-precision parallaxes. On one hand, an increasing amount of observational data of exquisite quality are now available, requiring reddening correction with similar or even higher precision (e.g., Xu et al. 2022). On the other hand, the intrinsic colors of stars can be predicted with a precision of approximately 1% from spectroscopic surveys, making it possible to achieve milli-magnitude-precision calibration of photometric surveys by the stellar color regression method (SCR; Yuan et al. 2015; Bowen et al. 2022). To compare predicted colors with observed colors and perform high-precision calibration with the SCR method, precise reddening correction is also required (e.g., Niu et al. 2021a, 2021b; Xiao & Yuan 2022; Huang & Yuan 2022). Reddening correction is becoming the critical factor that limits the optimal use of high-quality photometric, astrometric, and spectroscopic data. Hence, in the era of precision astronomy, millimagnitude-precision reddening correction is desired, from stellar physics to galactic archeology, extragalactic astronomy, and cosmology.

The whole-sky 2D dust-reddening map of Schlegel et al. (1998, hereafter SFD) has been widely used in reddening correction. Similar new reddening maps include those of Planck Collaboration et al. (2014) and Irfan et al. (2019). These 2D maps provide one of the best, if not the best, tools to perform reddening correction of Galactic and extragalactic targets in the middle- and high-Galactic-latitude regions.[3] The formulas $A(a) = R(a) \times E(B - V)$ and $E(a - b) = R(a - b) \times E(B - V)$ are usually used to estimate the extinction of a given passband $a$ and the color excess (reddening) of a given color $(a - b)$, respectively. Here, $R(a)$ and $R(a - b)$ are the extinction and reddening coefficients relative to $E(B - V)$, respectively. Their values can be either calculated theoretically from a given extinction curve or measured empirically using stars with known extinction and reddening.

However, there are still at least three problems to be solved before performing high-precision reddening correction with these maps, as follows. (1) The dust-emission-based 2D reddening maps suffer from a strong systematic spatial variation due to imperfect dust-emission modeling (see Sun et al. 2022 and references therein). (2) For a given extinction law, the reddening coefficient of a given color is not a constant value but a function of the extinction value and spectral energy distribution (SED; e.g., Jordi et al. 2010; Sale & Magorrian 2015; Danielski et al. 2018). (3) The extinction law of Galactic dust varies spatially (e.g., Schlafly et al. 2016, 2017) due to the variation in dust properties, such as dust-grain sizes and composition (Draine 2003).

---

[3] There has been significant development of dust maps in the last decade, particularly with the advent of 3D dust maps (e.g., Green et al. 2019; Chen et al. 2019). However, this paper does not attempt 3D mapping and focuses on the $T_{eff}$ and $E(B - V)$ dependence on the extinction law.

Thanks to millions of high-quality spectra and precise atmospheric parameters provided by the Large Sky Area Multi-Object Fiber Spectroscopic Telescope (LAMOST; Cui et al. 2012; Zhao et al. 2012; Liu et al. 2014), accurate reddening values in multiband colors can be determined toward millions of sight lines via the star-pair technique (Yuan et al. 2013), providing a great opportunity to address the aforementioned problems in unprecedented detail. To solve the first problem, using LAMOST Data Release (DR) 5 and Gaia DR2, Sun et al. (2022) have investigated the systematic error of the SFD and Planck extinction maps in the middle- and high-Galactic latitudes. Small yet significant spatially dependent biases are found for all the maps, which are related to the dust temperature, extinction, or dust spectral index. The results provide significant clues for the further improvement of the Galactic all-sky extinction maps.

In this work, we aim to investigate the second challenge by precisely measuring empirical reddening and extinction coefficients from the far-ultraviolet (UV) to the mid-infrared (IR). The measurement combines spectroscopic data from LAMOST with several widely used photometric surveys and considers the dependence on stellar SED and extinction. The spatial variations in the Galactic extinction law will be explored with the same data in a new paper. The effects arising from the three issues mentioned above might be coupled. More efforts are required to resolve the issues in a self-consistent way in the future.

The paper is organized as follows. In Section 2, we introduce the data sets for calculating reddening in different colors. A brief description of the star-pair method and the parameter setting is presented in Section 3. The reddening coefficients $R$ and $R(T_{eff}, E(B-V))$ are presented in Section 4, where the extinction coefficients for different passbands are also given. In Section 5, we compare the results with previous studies and model predictions. We summarize in Section 6.

## 2. Data

### 2.1. LAMOST

LAMOST is a 4 m quasi-meridian reflective Schmidt telescope with 4000 fibers that can concurrently observe up to 4000 objects each exposure in a field of view of 20 deg². LAMOST has collected over 9 million high-quality spectra during its first-stage low-resolution survey (LRS; 2011–2018, $R \sim 1800$ at 5500 Å). LAMOST has been conducting its second-stage survey (LAMOST II) since October 2018, which includes both low- and medium-resolution spectroscopic observations. In 2020 March, the LRS DR7 released 10,608,416 low-resolution spectra, covering the wavelength range 3690–9100 Å. The stellar parameters, including effective temperatures $T_{eff}$, surface gravities log $g$, metallicities [Fe/H], and radial velocities, have been derived from the LAMOST Stellar Parameter Pipeline for FGK stars, with a typical accuracy of about 150 K, 0.25 dex, 0.15 dex, and 5.0 km s⁻¹, respectively (Wu et al. 2011; Luo et al. 2015). The errors are dominated by systematic errors; the internal errors are much smaller (e.g., Niu et al. 2021a). Note that our method is concerned with the relative ranking of stars in the $T_{eff}$, log $g$, and [Fe/H] parameter spaces, and thus systematic uncertainties are not important. The systematic errors are independent of the extinction toward the stars, as the LAMOST stellar parameters are determined based on normalized spectra.

Since the LAMOST DR7 stellar parameter catalog does not contain hot stars, we complement it with a HotPayne catalog of Xiang et al. (2022). Using the PAYNE approach (Ting et al. 2019), this HotPayne catalog provides 11 stellar labels ($T_{eff}$, log $g$, projected rotation velocity $v_{sini}$, microturbulence velocity $v_{mic}$, $N_{He}/N_{tot}$, [C/H], [N/H], [O/H], [Si/H], [S/H], and [Fe/H]) of approximately 330,000 hot OBA stars with $T_{eff} \gtrsim 7500$ K. On the whole, the internal precision is approximately 2.1% for $T_{eff}$, 0.07 dex for log $g$, and 0.18 dex for [Fe/H]. Note that there are a significant number of metal-poor hot stars in the HotPayne catalog, which are mostly blue horizontal branch stars and blue straggler stars (Xiang et al. 2022). We crossmatch the stars in DR7 and HotPayne and attribute the intersection to the HotPayne catalog. The stellar parameters from the two catalogs are used for the star-pair algorithm, as well as the multiband photometric data to be introduced below.

### 2.2. GALEX

Galaxy Evolution Explorer (GALEX; Martin et al. 2005) is a space telescope that performs the first space UV sky survey in the far-UV (FUV; 1344–1786 Å) and near-UV (NUV; 1771–2831 Å) bands with 4″–6″ resolution. GALEX data play a significant role in analyzing blue and hot stars, as well as in characterizing dust properties.

### 2.3. SDSS

The Sloan Digital Sky Survey (SDSS) DR12 (Alam et al. 2015) provides accurate photometry of approximately 800 million stars in the $u_{sdss}$ (355 nm), $g_{sdss}$ (477 nm), $r_{sdss}$ (623 nm), $i_{sdss}$ (762 nm), and $z_{sdss}$ (913 nm) bands. In this work, we required $13 < u < 22.3$, $14 < g < 23.33$, $14 < r < 23.11$, $14 < i < 22.3$, and $12 < z < 20.8$ to avoid saturation (York et al. 2000).

### 2.4. Pan-STARRS 1

The Panoramic Survey Telescope and Rapid Response System 1 (Pan-STARRS 1, hereafter PS1; Chambers & Pan-STARRS Team 2018) 3π Steradian Survey is a multiepoch survey of the sky north of decl. $\delta = -30°$. The 3π survey was observed in five passbands $g_{PS}$, $r_{PS}$, $i_{PS}$, $z_{PS}$, and $y_{PS}$ spanning the range 400–1000 nm, with the mean 5σ point source limiting sensitivities of 23.3, 23.2, 23.1, 22.3, and 21.4 mag, respectively. To avoid saturation, for the $g/r/i/z/y_{PS}$ bands, we excluded stars brighter than 14.5, 14.5, 14.5, 14.0, and 13.0 mag, respectively.

### 2.5. Gaia

Gaia Early Data Release 3 (Gaia EDR3; Gaia Collaboration et al. 2021) delivers precise photometric data in three bands ($G$, BP, and RP) for over 1.8 billion sources acquired by the Gaia mission of the European Space Agency over its first 34 months of continuous all-sky scanning. Gaia EDR3 has provided the best photometric data to date, obtaining colors of unprecedented millimagnitude precision. However, modest magnitude-dependent calibration errors up to 10 mmag are found. Therefore, we adopt the magnitude corrections of Yang et al. (2021), which are based on the machine-learning technique and trained UBVRI magnitudes of approximately 10,000 Landolt standard stars into Gaia magnitudes. To exclude sources showing inconsistencies between the $G$-band, BP, and RP photometry, we adopt a limitation to the corrected BP and RP flux excess factor $C^*$: $|C^*| < 3\sigma_{C*}$, which is recommended by Riello et al. (2021, see their Section 6 for more details). Note that all distance data used in this work are from Gaia EDR3.

**Table 1**
Selection Criteria of Photometric Errors and $E(B-V)$ for the Control Samples of Different Passbands

| Passbands | Cuts on Photometric Errors | Criteria of $E(B-V)_{SFD}$ (Only for the LAMOST DR7 Catalog) |
|---|---|---|
| FUV | error(FUV) < 0.3 mag | $E(B-V) \leqslant 0.015$ mag; <br> if $T_{eff} > 6800$ K, $E(B-V) < 0.015 + (T_{eff} - 6800)/(5 \times 10^4)$ mag; <br> if $T_{eff} < 5500$ K, $E(B-V) < 0.015 - (T_{eff} - 5500)/(5 \times 10^4)$ mag. |
| NUV | error(NUV) < 0.3 mag | $E(B-V) \leqslant 0.01$ mag; <br> if [Fe/H] < −0.8, $E(B-V) < 0.01 - ([Fe/H] + 0.8)/100$ mag; <br> if $T_{eff} > 6200$ K, $E(B-V) < 0.01 + (T_{eff} - 6200)/(5 \times 10^4)$ mag; <br> if $T_{eff} < 5500$ K, $E(B-V) < 0.01 - (T_{eff} - 5500)/10^5$ mag. |
| SDSS passbands <br><br> PS1 passbands | $u/g/r/i/z_{sdss}$: 0.01, 0.005, 0.005, 0.006, 0.009 mag <br><br> $g/r/i/z/y_{PS}$: 0.008, 0.008, 0.006, 0.006, 0.008 mag | $E(B-V) \leqslant 0.01$ mag; <br> if [Fe/H] < −0.8, $E(B-V) < 0.01 - ([Fe/H] + 0.8)/100$ mag; <br> if $\log g < 4$, $E(B-V) < 0.01 - (\log g - 4)/80$ mag; <br> if $T_{eff} > 6200$ K, $E(B-V) < 0.01 + (T_{eff} - 6200)/(3 \times 10^4)$ mag; <br> if $T_{eff} < 5500$ K, $E(B-V) < 0.01 - (T_{eff} - 5500)/(3 \times 10^5)$ mag. |
| Gaia passbands <br> 2MASS passbands <br> W1, W2, and W3 | BP/RP: 0.004, 0.0045 mag (no limit for $G$ band) <br> $J/H/Ks$: 0.03, 0.04, 0.03 mag <br> W1/W2/W3: 0.03, 0.03, 0.15 mag | $E(B-V) \leqslant 0.01$ mag; <br> if [Fe/H] < −0.8, $E(B-V) < 0.01 - ([Fe/H] + 0.8)/100$ mag; <br> if $T_{eff} > 6200$ K, $E(B-V) < 0.01 + (T_{eff} - 6200)/(5 \times 10^4)$ mag; <br> if $T_{eff} < 5500$ K, $E(B-V) < 0.01 - (T_{eff} - 5500)/(3 \times 10^5)$ mag. |
| W4 | error(W4) < 0.3 mag | $E(B-V) \leqslant 0.02$ mag; <br> if $\log g > 3$, $E(B-V) < 0.02 + (\log g - 3)/40$ mag. |

### 2.6. 2MASS

The Two Micron All Sky Survey (2MASS; Skrutskie et al. 2006) has constructed a near-infrared $J$- (1.25 $\mu$m), $H$- (1.65 $\mu$m), and $Ks$-band (2.16 $\mu$m) image of the entire sky. The 2MASS Point Source Catalog contains photometry and astrometry for 471 million objects and achieves a $10\sigma$ point-source depth of $J = 15.8$, $H = 15.1$, and $Ks = 14.3$ mag.

### 2.7. WISE

The Wide-field Infrared Survey Explorer (WISE; Wright et al. 2010) is a mid-infrared survey of the entire sky, offering the four photometric bands W1, W2, W3, and W4 with wavelengths centered at 3.4, 4.6, 12, and 22 $\mu$m, respectively. Over 563 million pointlike and resolved objects are included in the WISE Source Catalog. Given the low sensitivities of the W3 and W4 bands, we adopted the following criteria to retain the bright stars with reliable photometry: W3 < 11.5; W4 < 8.4, and (W3 − W4) < 1.

### 3. Method

To acquire reddening values spanning from the far-UV to the mid-IR, we have implemented the star-pair algorithm (Yuan et al. 2013, see their Section 2.5 for more details). The approach is based on the assumption that stars with the same stellar atmosphere parameters have the same intrinsic colors. The intrinsic colors of a reddened target star may therefore be deduced from its control pairs/counterparts of the same atmospheric parameters that suffer from either nil or well-known extinction taken directly from the SFD map. The method requires carefully selected target and control samples to derive precise reddening values.

After filtering the magnitude ranges of the above surveys, we also use photometric error cuts to select the target sample: 0.3 mag for W3, W4, and GALEX passbands, 0.03 mag for $u_{sdss}$, 0.01 mag for PS1 and the remaining SDSS passbands, and 0.05 mag for W1, W2, and 2MASS passbands.

To ensure numerous stars and a relatively uniform distribution in the stellar parameter space of control samples, different detailed selection criteria are adopted for different passbands, where the criteria for photometric errors and reddening are listed in Table 1. For the reddening criteria, we first adopt a stringent limit; then, we loosen the limit for stars if they are in the relatively lower-density region of the stellar parameter space (for example, metal-poor stars). These $E(B-V)$ criteria apply only to the DR7 catalog. For the HotPayne catalog, we use the same reddening criteria for all the bands: $E(B-V) \leqslant 0.02$ mag; however, when $T_{eff} > 9000$ K, $E(B-V) < 0.02 + (T_{eff} - 9000)/70,000$ mag. The $E(B-V)$ limit is 0.22 mag when $T_{eff} = 23,000$ K

In addition to the photometric error cuts and reddening limits, we further require the following:

1. the vertical distance from the Galactic disk $|Z| > 300$ pc and an absolute value of ecliptic latitude greater than $10°$ to keep the control sample out of the Galactic and possible zodiacal dust. Note that the scale height of Galactic dust disk is approximately 103 pc (e.g., Li et al. 2018);

2. a signal-to-noise ratio (S/N) of the LAMOST spectra S/N > 20 to ensure reliable atmospheric parameters;

3. for the LAMOST DR7 catalog, error($T_{eff}$) < 300 K, error($\log g$) < 0.8 dex, and error([Fe/H]) < 0.4 dex. For the HotPayne catalog, due to the relatively small sample size, we loosened the error limits to error($T_{eff}$) < 600 K and error([Fe/H]) < 0.6 dex. We regard the S/N limits as the main way to control the quality of the stellar parameters, while the error limits of the stellar parameters are only an auxiliary way to remove the remaining few outliers.

Figure 1 shows the distributions of the target and control samples in the stellar parameter space, for four selected colors. The control samples cover the parameter space of the target samples. In Figures 1(b) and (c), the giant branches of the control stars show a double structure, corresponding to two sets of stars with different metallicities. The enhanced number of metal-poor stars in the control sample causes the presence of the upper branch. In addition, there are very few red clumps in panel (b) compared to the red clumps in panel (c) because red-clump stars in the high-Galactic-latitude regions are so bright that they are saturated in the PS1 data. The metal-poor hot stars from the HotPayne catalog in panel (d) are mostly blue horizontal branch stars and blue straggler stars (Xiang et al. 2022).

For each star in the target samples, its control stars are selected from the control sample as their $T_{eff}$, $\log g$, and [Fe/H] differ from those of the target by satisfying $\Delta T_{eff} < 30 + [0.005 \times (6000 - T_{eff})]^2$ K, $\Delta \log g < 0.5$ dex, and $\Delta$[Fe/H] $< 0.3$ dex, respectively. Then, to determine the intrinsic color of the target star, a simple linear fit to the local control stars is performed: $\text{color}_0 = a \times T_{eff} + b \times \log g + c \times$ [Fe/H] $+ d$, where $a$, $b$, $c$, and $d$ are free parameters. To guarantee the fitting quality, we also put a lower limit on the number of selected control stars: 5 for (NUV $- g_{PS}$), 10 for optical colors, 15 for (W2 $-$ W3), and 30 for the other infrared colors.

For each star in the control samples, its intrinsic colors can be estimated in two independent ways. One is based on its stellar spectroscopic parameters and estimated using the star-pair algorithm; the target stars can include all stars, i.e., the control stars can also be target stars. (named color$_{0sp}$). The other way is based on the SFD map and simply estimated as $\text{color}_0 = \text{color}_{obs} - R_{color}(T_{eff}, E(B-V)) \times E(B-V)$. $R_{Color}(T_{eff}, E(B-V))$ is obtained iteratively in this work.

We calculate the differences between the two measurements (color$_0 -$ color$_{0sp}$), of which the distributions are well fitted by a Gaussian in most cases (except W2 $-$ W3). The mean and standard deviation (std) values are compared and displayed in Figure 2. Then, the 3-std (2-std for colors involving W1 or W2) outliers are iteratively excluded from the control samples until convergence. The std values indicate the reddening accuracies achieved with the star-pair algorithm. The typical standard deviation varies in different colors, e.g., $<0.01$ mag for Gaia colors, $\sim0.01$ mag for PS1 colors, and $\sim 0.02$–$0.03$ mag for SDSS and 2MASS colors. The final control and target sample sizes of the DR7 and HotPayne catalogs are shown in Figure 3.

## 4. The FUV to MIR Reddening Coefficients $R$ and $R(T_{eff}, E(B-V))$

### 4.1. Single-valued Reddening Coefficients $R$

For a given color $(a - b)$, the reddening is calculated by $E(a - b) = (a - b)_{obs} - (a - b)_0$, where $(a - b)_{obs}$ is the observed color and $(a - b)_0$ is the predicted intrinsic color. The reddening coefficient $R$ is defined by $R(a - b) = E(a - b)/E(B - V)_{SFD}$. To calculate the single-valued $R$ for each color, we first combine the reddening results of the DR7 and HotPayne catalogs. Then, further sample selection is performed with the criteria below:

1. $|Z| > 300$ pc and Galactic latitude $|b| > 10°$;
2. $|$ ecliptic latitude $| > 10°$;
3. S/N $> 10$;
4. $E(B - V)_{bayes19} > 0.85 \times E(B - V)_{SFD} - 0.2$, where $E(B - V)_{bayes19}$ is interpolated from the 3D extinction map of Green et al. (2019). This empirical criterion, obtained from the comparison of two sources of $E(B - V)$, is to further exclude stars of overestimated $E(B - V)_{SFD}$ in the low-Galactic-latitude regions.

For Gaia passbands, we make a finer rejection of stars suffering background and nearby star contamination issues (Riello et al. 2021). First, by using the extinction correction results of Niu et al. (2021b), we obtain the dereddened magnitudes and intrinsic color (BP $-$ RP)$_0$. Then, combined with the zero-points of Gaia EDR3 passbands, we calculate the excess factor by (Flux$_{BP}$ + Flux$_{RP}$)/Flux$_G$. We finally require the same empirical cut in Xu et al. (2022): excess factor $< 0.091 \times$ (BP $-$ RP)$_0 + 1.15$, which eliminates 13.5% of sources in Gaia colors.

For selected samples, we equally divide $E(B - V)_{SFD}$ into 10 bins from 0 to 0.5 mag, then the medians of $E(B - V)$ and $E(a - b)$ are obtained after $3\sigma$ clipping. The error of median $E(a - b)$ for a given bin is estimated by std$(E(a - b))/\sqrt{N}$, where $N$ is the number of sources in the bin. A weighted linear fitting to the medians is adopted to determine the reddening coefficients $R$, using the inverse of the squared error as the weight. The fitting results for the 21 colors are plotted in Figure 4 and listed in Table 2.

For most colors, the intercepts of the linear function are less than 0.005 mag, which implies that the measurements of the star-pair method match the SFD extinction for low-extinction sources. The intercept for the (FUV–NUV) color is $-0.021$ mag, probably due to the large photometric errors and small number of sample stars in the FUV band. The same reason also causes a wider scatter distribution of the (W3 – W4) color in Figure 4.

As seen in Figure 4, the medians show downward deviations from the fitted lines as $E(B - V)$ increases in almost all colors, which is stronger in the UV and optical and weaker in the IR, suggesting that the reddening coefficients $R$ are $E(B - V)$ dependent due to the nonlinear response of extinction with increasing dust column.[4] There is also a clear $T_{eff}$ dependence in the (NUV $- g_{PS}$), (BP $- G$), and ($G -$ RP) colors, which is not surprising because the stellar extinction in a given passband is determined by the convolution of the stellar SED with the passband throughput and the extinction curve. As a result, for a given extinction curve, reddening coefficients $R$ generally depend on stellar SED ($T_{eff}$ for simplicity[5]) and extinction $E(B - V)$. The broader filters (e.g., the Gaia passbands) or the bluer filters (e.g., the GALEX filters), the stronger the dependence. Empirical temperature- and reddening-dependent reddening coefficients have been provided for the Gaia DR2 (Sun et al. 2022) and EDR3 colors (Niu et al. 2021b). Empirical color- and reddening-dependent reddening coefficients are also provided for the Gaia EDR3 colors (Xu et al. 2022).

---

[4]  Note that Schlafly et al. (2016) and our work in progress have shown that the extinction law is uncorrelated with $E(B - V)$. Therefore, the possibility that the dependence derived on $E(B - V)$ comes from a variation in the extinction law in regions of higher extinction is very low.

[5]  The effect of metallicity and surface gravity on stellar SED is much weaker than effective temperature.

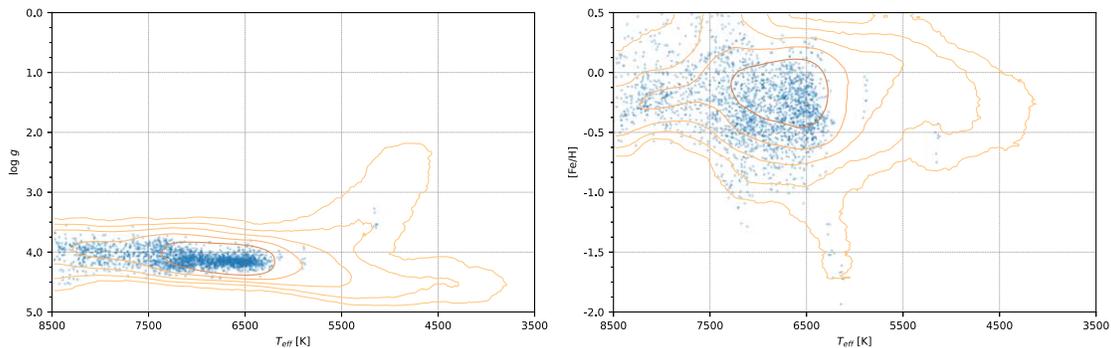

(a) LAMOST DR7 ($FUV - NUV$) color

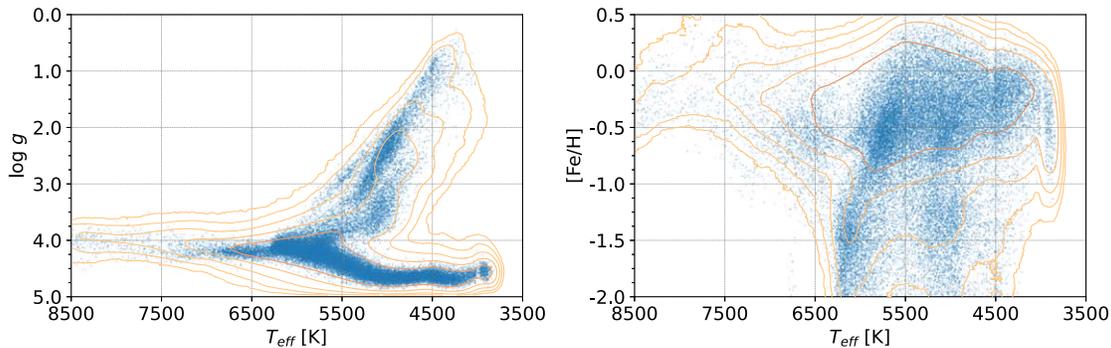

(b) LAMOST DR7 ($g_{PS} - r_{PS}$) color

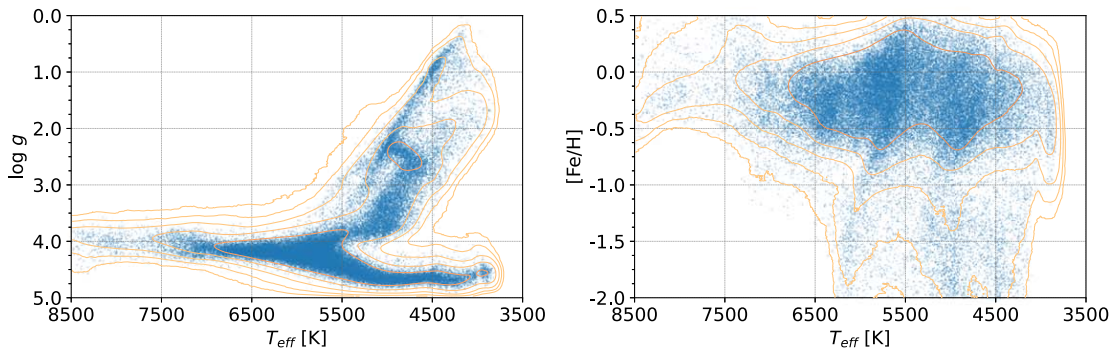

(c) LAMOST DR7 ($H - Ks$) color

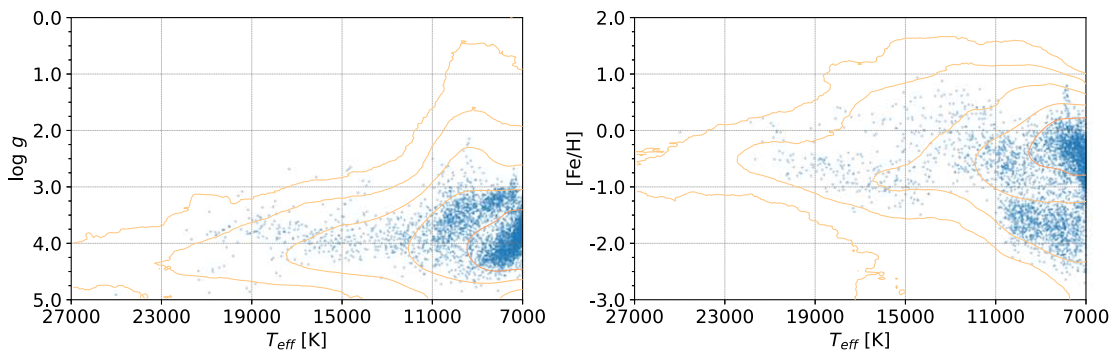

(d) HotPayne catalog ($BP - G$) color

**Figure 1.** Examples of $T_{\mathrm{eff}}$ vs. $\log g$ (left panels) and $T_{\mathrm{eff}}$ versus [Fe/H] (right panels) distributions of the target and control samples. From top to bottom: FUV–NUV, $g_{\mathrm{PS}} - r_{\mathrm{PS}}$, $H - Ks$, and BP – $G$ colors, respectively. In each panel, the blue dots are the control sample stars, and the density contour represents the target sample of the given color.

**Table 2**
Single-valued Reddening Coefficients $R$ and Fitting Coefficients of $R(T_{\rm eff}, E(B-V))$ for 21 Colors

| Color | $R$ | $C_1$ | $C_2$ | $C_3$ | $C_4$ | $C_5$ | $C_6$ |
|---|---|---|---|---|---|---|---|
| $FUV-NUV$ | $-0.319$ | $4.65e{-}10$ | $-1.21e{-}05$ | $0.104$ | $7.541$ | $-6.793$ | $-300.238$ |
| $NUV-g_{PS}$ | $4.045$ | $9.23e{-}11$ | $-2.24e{-}06$ | $0.0184$ | $-0.466$ | $-2.896$ | $-45.589$ |
| $g_{PS}-r_{PS}$ | $0.885$ | $-3.22e{-}11$ | $6.09e{-}07$ | $-0.00378$ | $-0.284$ | $-0.170$ | $8.630$ |
| $r_{PS}-i_{PS}$ | $0.572$ | $-2.2e{-}11$ | $4.19e{-}07$ | $-0.00262$ | $0.145$ | $-0.292$ | $6.006$ |
| $i_{PS}-z_{PS}$ | $0.392$ | $-4.35e{-}12$ | $9.13e{-}08$ | $-0.000622$ | $0.458$ | $-0.503$ | $1.877$ |
| $z_{PS}-y_{PS}$ | $0.252$ | $-4.58e{-}12$ | $9.46e{-}08$ | $-0.000635$ | $0.081$ | $-0.154$ | $1.673$ |
| $y_{PS}-J$ | $0.399$ | $-2.27e{-}11$ | $4.32e{-}07$ | $-0.00272$ | $-1.012$ | $0.583$ | $5.979$ |
| $J-H$ | $0.295$ | $-4.38e{-}12$ | $8.27e{-}08$ | $-0.000527$ | $0.421$ | $-0.401$ | $1.506$ |
| $H-Ks$ | $0.147$ | $3.85e{-}13$ | $-9.67e{-}09$ | $7.91e{-}05$ | $-0.366$ | $0.225$ | $-0.093$ |
| $Ks-W1$ | $0.112$ | $7.18e{-}13$ | $-1.11e{-}08$ | $5.18e{-}05$ | $-0.128$ | $0.038$ | $0.045$ |
| $W1-W2$ | $0.056$ | $-6.11e{-}12$ | $1.11e{-}07$ | $-0.000667$ | $0.098$ | $-0.099$ | $1.411$ |
| $W2-W3$ | $-0.045$ | $\cdots$ | $\cdots$ | $\cdots$ | $\cdots$ | $\cdots$ | $\cdots$ |
| $W3-W4$ | $0.099$ | $\cdots$ | $\cdots$ | $\cdots$ | $\cdots$ | $\cdots$ | $\cdots$ |
| $BP-G$ | $0.634$ | $-2.19e{-}12$ | $5.21e{-}08$ | $-0.00043$ | $0.008$ | $-0.150$ | $1.809$ |
| $G-RP$ | $0.627$ | $-2.67e{-}12$ | $4.85e{-}08$ | $-0.00019$ | $-0.051$ | $-0.324$ | $0.706$ |
| $RP-Ks$ | $1.431$ | $-7.88e{-}12$ | $1.76e{-}07$ | $-0.00126$ | $-0.630$ | $-0.057$ | $4.364$ |
| $u_{sdss}-g_{sdss}$ | $1.048$ | $-3.41e{-}11$ | $6.96e{-}07$ | $-0.00465$ | $1.520$ | $-1.550$ | $11.508$ |
| $g_{sdss}-r_{sdss}$ | $1.052$ | $-3.99e{-}11$ | $7.47e{-}07$ | $-0.00459$ | $0.301$ | $-0.642$ | $10.443$ |
| $r_{sdss}-i_{sdss}$ | $0.601$ | $-1.49e{-}11$ | $2.93e{-}07$ | $-0.00188$ | $0.595$ | $-0.588$ | $4.678$ |
| $i_{sdss}-z_{sdss}$ | $0.500$ | $-4.86e{-}12$ | $1.01e{-}07$ | $-0.000694$ | $1.353$ | $-1.124$ | $2.252$ |
| $z_{sdss}-Ks$ | $0.993$ | $-3.49e{-}11$ | $6.79e{-}07$ | $-0.00433$ | $-2.027$ | $1.144$ | $9.892$ |

**Note.** The function form is $R(T_{\rm eff}, E(B-V)) = C_1 \times T_{\rm eff}^3 + C_2 \times T_{\rm eff}^2 + C_3 \times T_{\rm eff} + C_4 \times E(B-V)^2 + C_5 \times E(B-V) + C_6$.

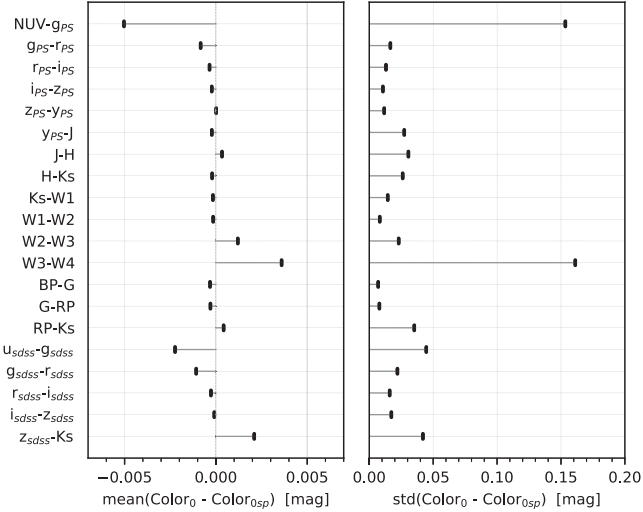

**Figure 2.** Mean and standard deviation of the differences between the two independent measurements of intrinsic colors (color$_0$–color$_{0sp}$). Note the values for the (FUV − NUV) color, −0.052 and 0.306 respectively, are too large to be shown.

### 4.2. Temperature- and Reddening-dependent Reddening Coefficients $R(T_{\rm eff}, E(B-V))$

To precisely portray the reddening coefficients $R(T_{\rm eff}, E(B-V))$ as a function of $T_{\rm eff}$ and $E(B-V)$, we divide the selected target samples into two-dimensional grids of $T_{\rm eff}$ and $E(B-V)$. $T_{\rm eff}$ values are first divided into six equal bins from 4000 to 10,000 K. Considering that there are many sources between 4000 and 5000 K, this range is further divided into two equal bins. $E(B-V)$ values are divided into 10 equal bins from 0.0 to 0.5 mag. We discard the 0–0.05 bin, as the small reddening in this bin can cause large errors in the calculation of $R$. For each grid, the medians of $R(T_{\rm eff}, E(B-V))$, $E(B-V)$, and $T_{\rm eff}$ are calculated after $3\sigma$ clipping. The errors of $R(T_{\rm eff}, E(B-V))$ are first estimated by error$(R) = {\rm std}(R)/\sqrt{N}$. Then, the median of the errors is added to all error values to mitigate their differences due to large variations in $N$. After discarding a few grids with fewer than 20 stars, we carry out a weighted polynomial fit for the grid medians. Here, we use the following binary function:

$$R(T_{\rm eff}, E(B-V)) = C_1 \times T_{\rm eff}^3 + C_2 \times T_{\rm eff}^2 + C_3 \times T_{\rm eff}$$
$$+ C_4 \times E(B-V)^2 + C_5 \times E(B-V) + C_6.$$

No cross terms are needed here based on the data. This result is further confirmed by the subsequent model results in Section 5.1. The fitting results and residuals are shown in Figures 5 and 6, respectively. The fitting coefficients are listed in Table 2. For the (W2 − W3) and (W3 − W4) colors, fitting is not performed due to their large errors.

As seen from Figure 5, the reddening coefficients decrease with $E(B-V)$ for most colors, except for the $(y_{PS} - J)$, $(H - Ks)$, and $(z_{sdss} - Ks)$ colors, which first increase and then decrease. The increase at the low-extinction region $(E(B-V) < 0.3)$ of the three colors is probably caused by their negative intercepts in Figure 4. For most colors, there is also a clear $T_{\rm eff}$ dependence, particularly for the (FUV–NUV), (NUV − $g_{PS}$), (BP − $G$), and (G − RP) colors. Usually, the reddening coefficients vary (increase in most cases) with $T_{\rm eff}$ monotonically. The monotonically increasing nature of $R$ with $T_{\rm eff}$ for the PS1 and SDSS colors is broken at the low-temperature end (4000–5500 K, see the red and orange curves in Figure 5). For the two optical surveys mentioned above, the spatial distributions of the low-temperature stars and hotter stars are different, with the latter being more concentrated in the low-Galactic-latitude region ($|b| < 30°$). Therefore, we speculate that it may be related to the spatial variations in $R(V)$

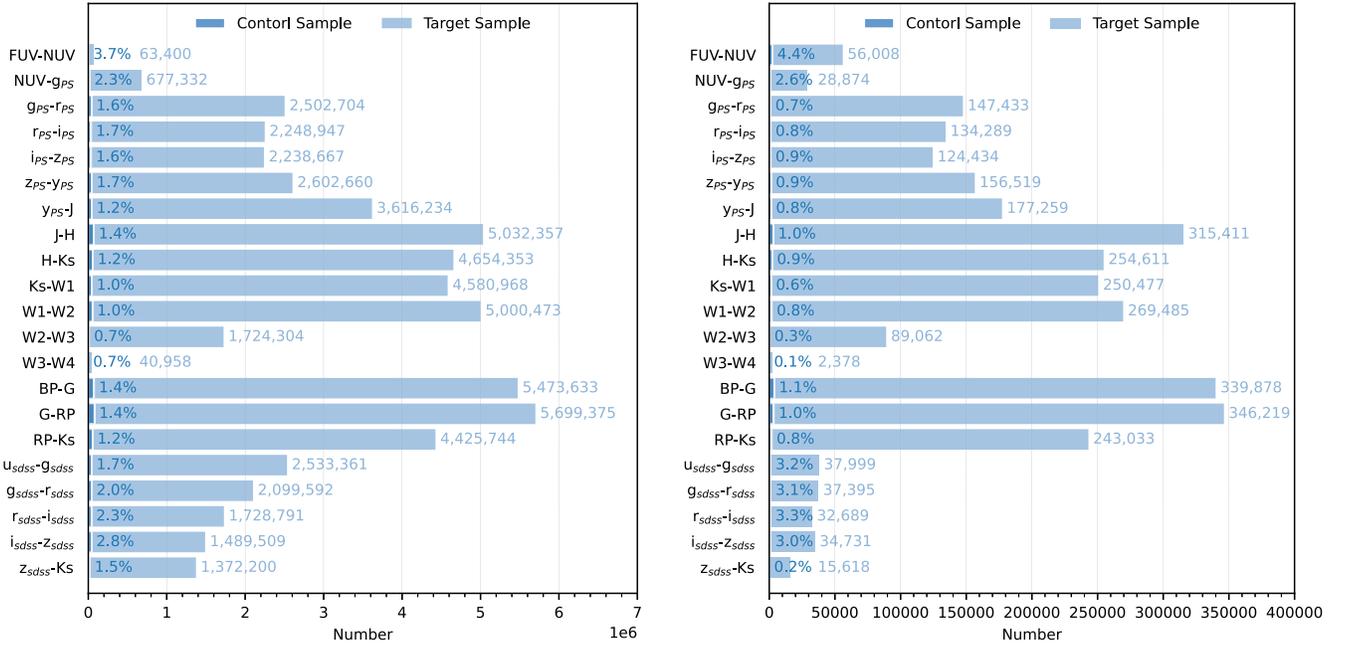

**Figure 3.** The star numbers of control and target samples of different colors for the LAMOST DR7 (left panel) and HotPayne (right panel) catalogs. The relative fractions, with a typical value of 1%, and total numbers of the target samples are marked.

**Table 3**
Single-valued Extinction Coefficients $R$ and Fitting Coefficients of $R(T_{eff}, E(B-V))$ for 22 Bands

| Bands | $R$ | $C_1$ | $C_2$ | $C_3$ | $C_4$ | $C_5$ | $C_6$ |
|---|---|---|---|---|---|---|---|
| FUV | 6.973 | $4.68e-10$ | $-1.26e-05$ | 0.112 | 6.520 | $-10.401$ | $-319.943$ |
| NUV | 7.293 | $2.57e-12$ | $-5.19e-07$ | 0.0076 | $-1.022$ | $-3.608$ | $-19.704$ |
| $g_{PS}$ | 3.248 | $-8.98e-11$ | $1.72e-06$ | $-0.0108$ | $-0.556$ | $-0.712$ | 25.885 |
| $r_{PS}$ | 2.363 | $-5.76e-11$ | $1.11e-06$ | $-0.00704$ | $-0.273$ | $-0.542$ | 17.255 |
| $i_{PS}$ | 1.791 | $-3.56e-11$ | $6.91e-07$ | $-0.00443$ | $-0.417$ | $-0.250$ | 11.249 |
| $z_{PS}$ | 1.398 | $-3.12e-11$ | $6.00e-07$ | $-0.00381$ | $-0.876$ | 0.253 | 9.372 |
| $y_{PS}$ | 1.146 | $-2.67e-11$ | $5.05e-07$ | $-0.00317$ | $-0.957$ | 0.407 | 7.699 |
| $J$ | 0.748 | $-3.99e-12$ | $7.30e-08$ | $-0.000448$ | 0.055 | $-0.176$ | 1.720 |
| $H$ | 0.453 | $3.85e-13$ | $-9.67e-09$ | $7.91e-05$ | $-0.366$ | 0.225 | 0.213 |
| $Ks$ | 0.306 | $\cdots$ | $\cdots$ | $\cdots$ | $\cdots$ | $\cdots$ | $\cdots$ |
| W1 | 0.194 | $-7.18e-13$ | $1.11e-08$ | $-5.18e-05$ | 0.128 | $-0.038$ | 0.261 |
| W2 | 0.138 | $5.40e-12$ | $-9.97e-08$ | 0.000615 | 0.030 | 0.062 | $-1.149$ |
| W3 | 0.183 | $\cdots$ | $\cdots$ | $\cdots$ | $\cdots$ | $\cdots$ | $\cdots$ |
| W4 | 0.084 | $\cdots$ | $\cdots$ | $\cdots$ | $\cdots$ | $\cdots$ | $\cdots$ |
| BP | 2.998 | $-1.27e-11$ | $2.76e-07$ | $-0.00188$ | $-0.673$ | $-0.531$ | 7.185 |
| $G$ | 2.364 | $-1.05e-11$ | $2.24e-07$ | $-0.00145$ | $-0.681$ | $-0.381$ | 5.376 |
| RP | 1.737 | $-7.88e-12$ | $1.76e-07$ | $-0.00126$ | $-0.630$ | $-0.057$ | 4.670 |
| $u_{sdss}$ | 4.500 | $-1.29e-10$ | $2.52e-06$ | $-0.0162$ | 1.741 | $-2.760$ | 39.080 |
| $g_{sdss}$ | 3.452 | $-9.46e-11$ | $1.82e-06$ | $-0.0115$ | 0.221 | $-1.210$ | 27.571 |
| $r_{sdss}$ | 2.400 | $-5.47e-11$ | $1.07e-06$ | $-0.00691$ | $-0.080$ | $-0.568$ | 17.129 |
| $i_{sdss}$ | 1.799 | $-3.98e-11$ | $7.80e-07$ | $-0.00502$ | $-0.675$ | 0.020 | 12.450 |
| $z_{sdss}$ | 1.299 | $-3.49e-11$ | $6.79e-07$ | $-0.00433$ | $-2.027$ | 1.144 | 10.198 |

**Note.** The function form is $R(T_{eff}, E(B-V)) = C_1 \times T_{eff}^3 + C_2 \times T_{eff}^2 + C_3 \times T_{eff} + C_4 \times E(B-V)^2 + C_5 \times E(B-V) + C_6$.

and/or the spatially dependent systematic SFD map. Further investigations will be performed in the future.

### 4.3. Single-valued and Temperature- and Reddening-dependent Extinction Coefficients

To derive the extinction coefficients from the reddening coefficients presented above, we adopt the reference point $R(Ks) = 0.306$ estimated by Yuan et al. (2013), which is predicted by the $R(V) = 3.1$ Fitzpatrick law assuming $E(B-V)_{SFD}$ overpredicted by 14% (Schlafly & Finkbeiner 2011; Yuan et al. 2013). The resultant single-valued extinction coefficients and polynomial coefficients are given in Table 3. Reddening coefficients of other combinations of colors can be estimated accordingly.

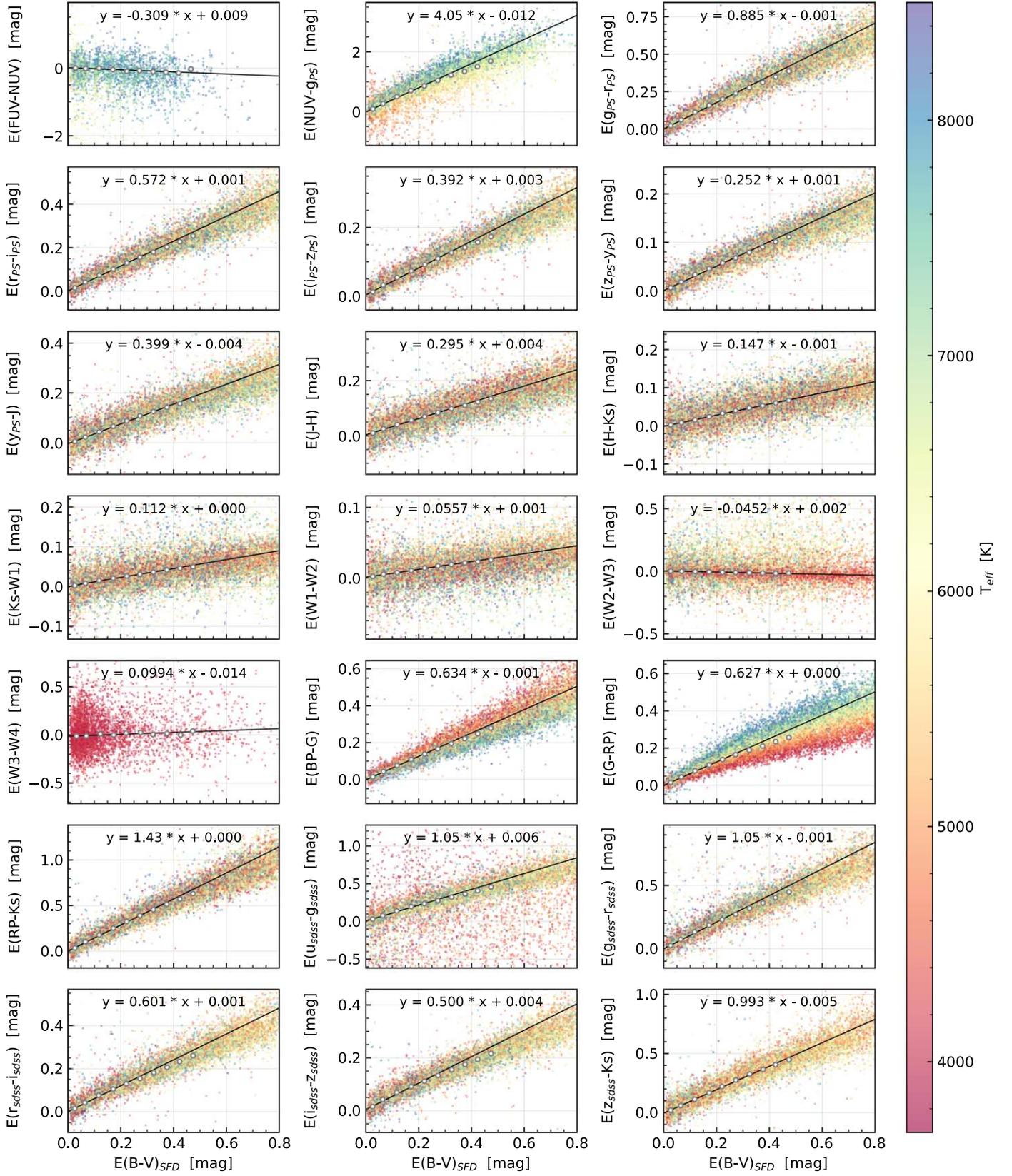

**Figure 4.** The linear fitting results of reddening coefficients in 21 colors. The white circles denote the median values by binning the selected target samples into 10 groups with a bin size of 0.05 in the *x*-axis, and the statistical errors are too small to be shown. The black lines represent weighted linear regression of the white circles. Note that we plot as equal number of stars as possible in each temperature and $E(B-V)$ interval to clearly show that $R$ is $T_{\mathrm{eff}}$ dependent.

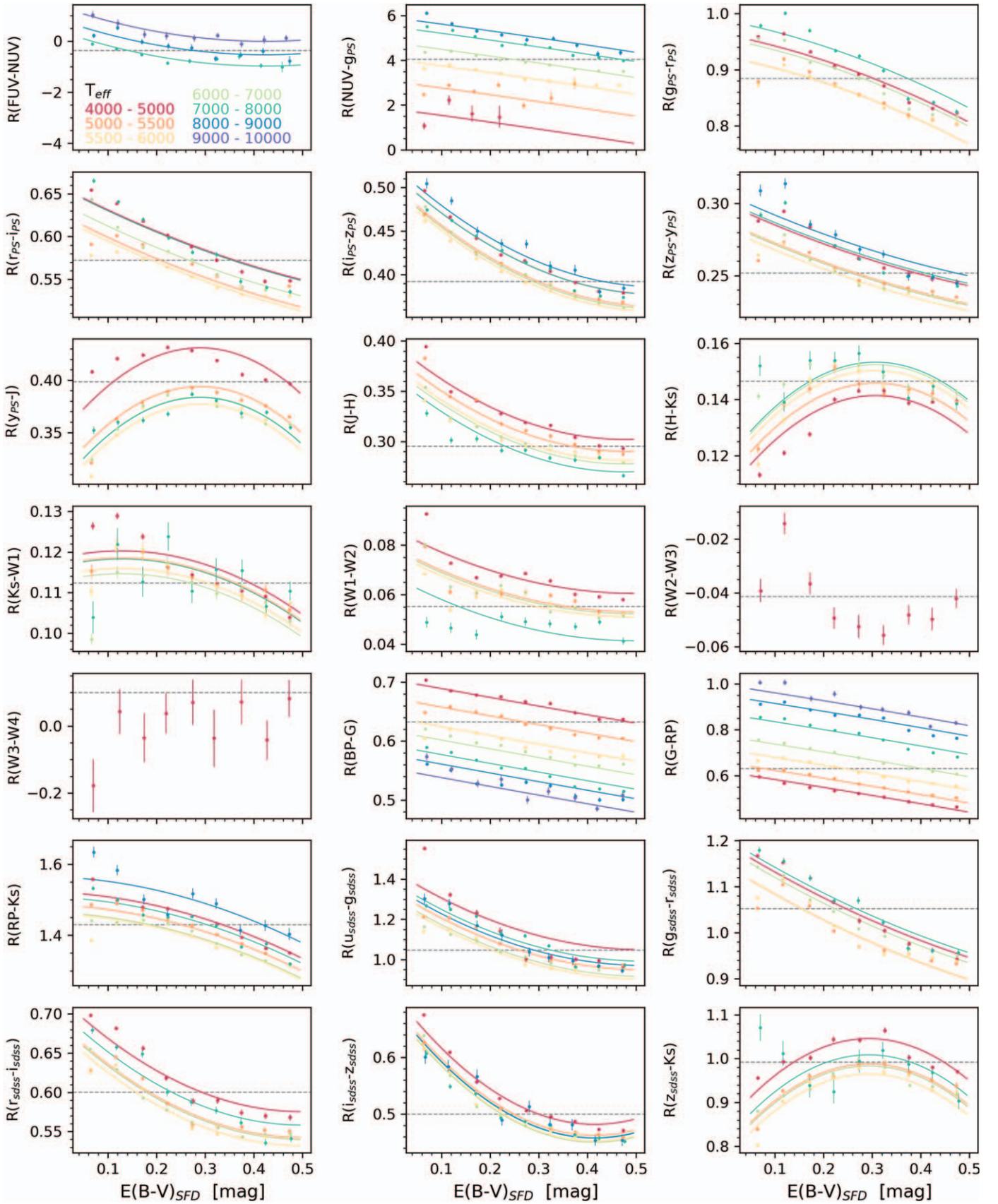

**Figure 5.** Reddening coefficients as a function of $T_{eff}$ and $E(B - V)$ of 21 colors. The color-coded points are the median values of the gridded target sample, with colors denoting $T_{eff}$ bins. The error bars are overplotted. The corresponding colored curves represent the fitting results. The black dashed lines show the single-valued reddening coefficients $R$.

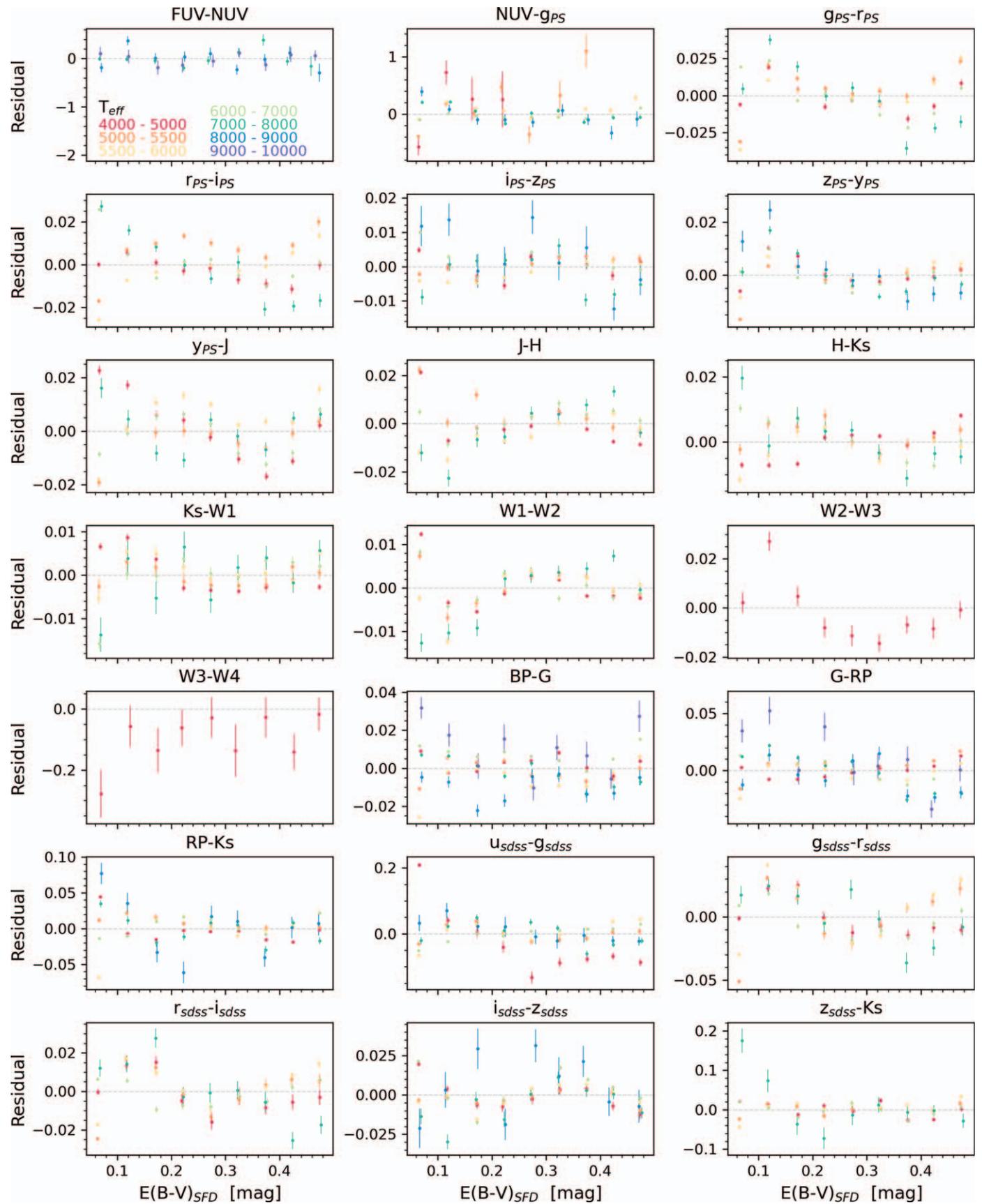

**Figure 6.** Same as Figure 5 but for the fitting residuals.

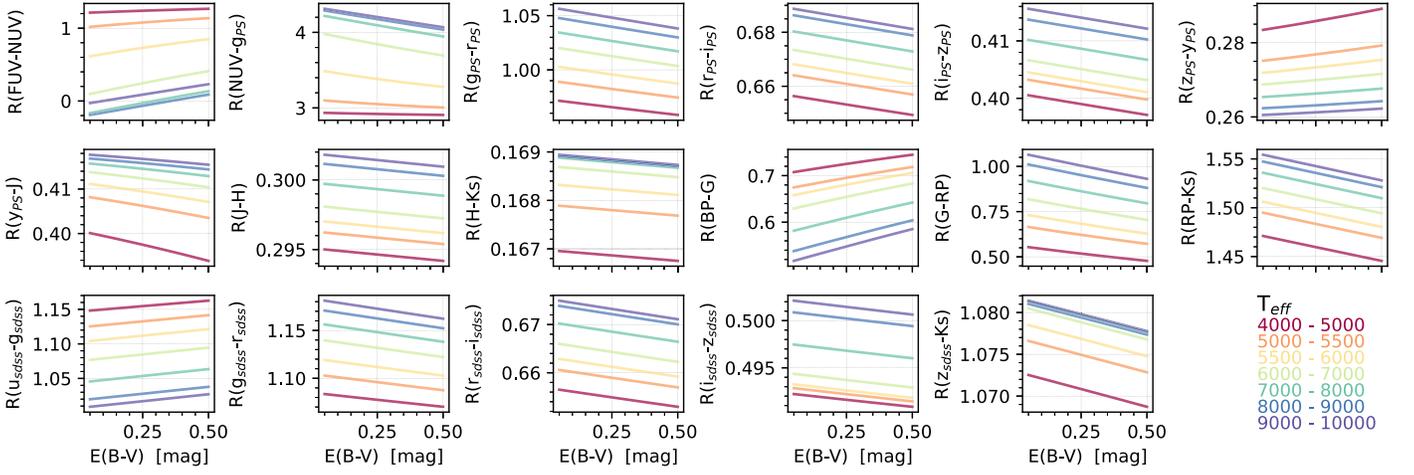

(a) Temperature- and reddening-dependent reddening coefficients predicted by the F99 extinction law of R(V) = 3.1.

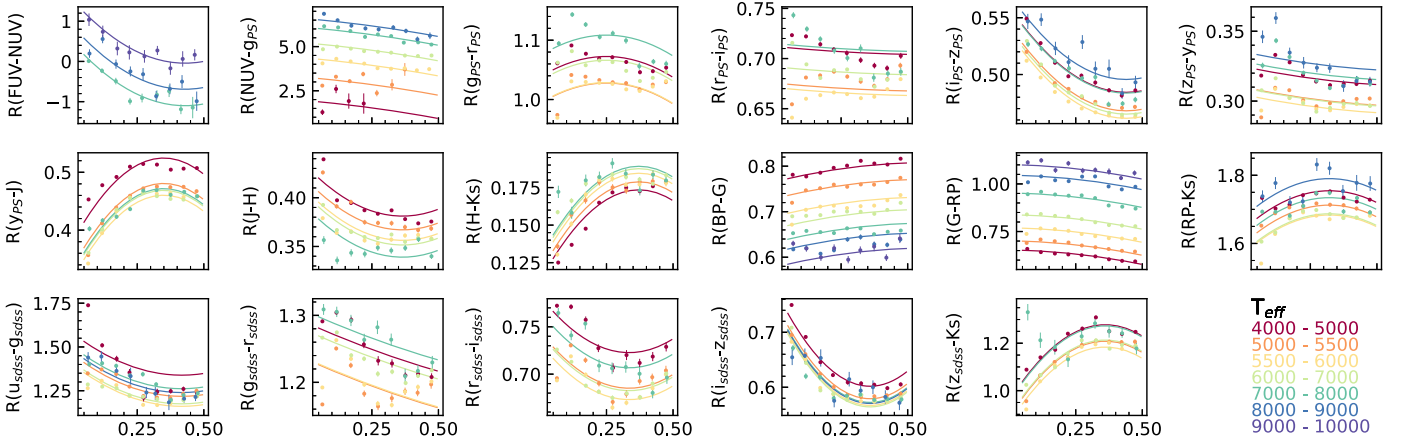

(b) Empirical reddening coefficients with $E(B-V)_{SFD}$ corrected (see text for more detail).

**Figure 7.** The comparison of the predicted (a) and corrected empirical functions (b) of reddening coefficients $R(T_{eff}, E(B-V))$ for 17 colors. The symbols and lines are similar to those in Figure 5.

## 5. Discussion

### 5.1. Comparison with Model Results

In this subsection, we first compute the model-predicted reddening coefficients $R(T_{eff}, E(B-V))$ and then compare them with the observed ones. For stellar SEDs, we use the "BOSZ" stellar atmosphere models (Bohlin et al. 2017). For the reddening laws, we adopt the $R(V)$-dependent reddening law of Fitzpatrick (1999, hereafter F99). The F99 reddening law of $R(V) = 3.1$ is used because it matches observations (e.g., Schlafly & Finkbeiner 2011; Yuan et al. 2013). We select the BOSZ model spectra with stellar parameters $[M/H] = 0$, $[\alpha/H] = 0$, $[C/H] = 0$, and $\log g = 4$. We have tried different $\log g$ and found that the effect on the results is too small to have an effect. Combining the filter profiles of each survey (Rodrigo & Solano 2020), we calculate the reddening coefficients for stars of different temperatures and extinction values. Since the reddest wavelength of the BOSZ spectra is 3 $\mu$m, the four WISE bands are excluded in the calculation. The predicted reddening coefficients are displayed in Figure 7(a).

The SFD map overestimates reddening in a reddening-dependent manner (Sun et al. 2022). To correct such an effect, we compare the empirical and predicted results of the $r_{PS} - i_{PS}$ color for the 5500–6000 K temperature interval. We force them to match by correcting $E(B-V)_{SFD}$ with a quadratic function. All empirical reddening coefficients are revised accordingly and plotted in Figure 7(b).

The predicted and empirical reddening coefficients generally match, particularly for optical colors. The agreement is excellent for the two Gaia colors. The agreement is reasonably good for most PS1 and SDSS colors, if the aforementioned nonmonotonic problem is ignored at the low-temperature end. However, for the $(z_{PS} - y_{PS})$ and $(u_{sdss} - g_{sdss})$ colors, the temperature trends are opposite. Differently from most colors, the model predicts higher reddening coefficients at lower temperatures for $(z_{PS} - y_{PS})$ and $(u_{sdss} - g_{sdss})$ colors, probably due to the effect of Balmer and Paschen jumps. However, we observe the opposite trend. The consistency becomes slightly worse for infrared colors. The temperature trend is opposite for the $(J - H)$ color, and the observed dependence on temperature is much stronger than predicted. The prediction of the (FUV–NUV) color is not consistent with ours, which may be due to the possible problems of the stellar atmosphere models as well as the F99 extinction curve in the far-UV.

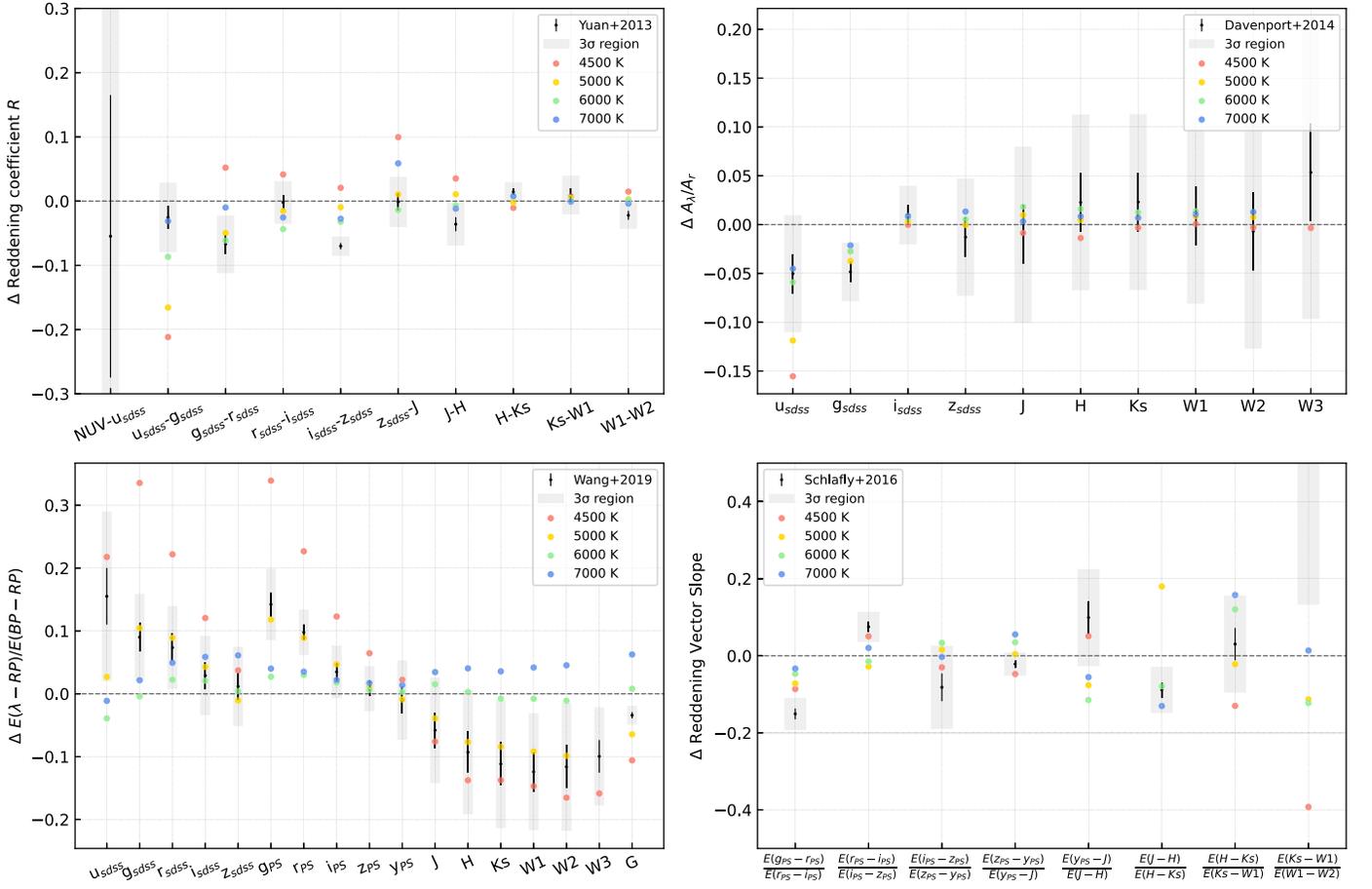

**Figure 8.** Comparison with the results in the literature. The black dots with error bars show the literature values and errors. The gray bars represent the corresponding $3\sigma$ range. The colored dots represent the calculated values of the function at different $T_{\rm eff}$.

### 5.2. Comparison with the Literature

Reddening/extinction coefficients of various photometric colors/bands have been widely determined. However, their variations with stellar SED and extinction are rarely depicted. In this subsection, we compare our results with the results reported in the literature. To facilitate comparisons, we adopt $R(T_{\rm eff}, E(B-V))$ values at $T_{\rm eff} = 4500/5000/6000/7000$ K and $E(B-V) = 0.5$ mag, except for those with Yuan et al. (2013), where we adopt $E(B-V) = 0.3$ mag because most sample stars of Yuan et al. (2013) are in the high-Galactic-latitude region and are of small extinction. Then, we convert our results to the form in the literature. In Figure 8, both the $R(T_{\rm eff}, E(B-V))$-based results and the literature results are compared with those based on our single-valued reddening coefficients.

Using the star-pair technique, Yuan et al. (2013) measured dust-reddening coefficients for the GALEX, SDSS, 2MASS, and WISE passbands. The single-valued reddening coefficients in the current work are consistent with the literature values, except for the $(i_{\rm sdss} - z_{\rm sdss})$ color. The typical best-match temperature is between 6000 and 7000 K, and varies between different colors.

Davenport et al. (2014) measured the dust extinction curve in the SDSS–2MASS–WISE bands relative to the $r_{\rm sdss}$ band by the SDSS–2MASS–WISE 10-dimensional stellar color locus. The upper-right panel of Figure 8 shows that our results at each temperature and the literature values are consistent with the uncertainties. The results are most consistent at 6000 K, when all bands are at the $1\sigma$ level except for $g_{\rm sdss}$, which is at the $2\sigma$ level. Davenport et al. (2014) reported variations in infrared dust extinction with both Galactic latitude and longitude, which might be related to different samples used.

Wang & Chen (2019) obtained the optical–mid-IR extinction $A_\lambda/A_{RP}$ in a total of 21 bands, by using spectroscopic data from APOGEE red-clump stars and multiband photometric data from Gaia, APASS, SDSS, PS1, 2MASS, and WISE, as shown in the bottom-left panel of Figure 8. The agreement between Wang & Chen (2019) and our calculations at 5000 K is excellent, within $1\sigma$ in almost all the jointly studied bands. The APOGEE red-clump stars of Wang & Chen (2019) have a typical $T_{\rm eff}$ of 4900 K, which explains why their results match best with ours at 5000 K.

Schlafly et al. (2016) calculated multiband color-excess ratios by using spectroscopic data from the APOGEE (with typical $E(B-V)$ of 0.65 mag and typical $T_{\rm eff}$ between 4000–5000 K) and 10-band photometry from PS1, 2MASS, and WISE. The results are compared in the bottom-right panel of Figure 8. Our results at $T_{\rm eff} = 4500$ K match theirs within $2\sigma$ in most cases. However, there is a large discrepancy in the color-excess ratios $E(J-H)/E(H-Ks)$ and $E(Ks-W1)/E$ $(W1 - W2)$. We obtain $E(J - H)/E(H - Ks) = 2.292$ and $E(Ks - W1)/E(W1 - W2) = 1.534$ at $T_{eff} = 4500$ K, while their values are 1.943 and 2.627, respectively.

Reddening coefficients in different works are not always consistent with each other. One important reason is that they are derived from samples at different temperatures and extinction distributions. The empirical temperature- and reddening-dependent coefficients $R(T_{eff}, E(B - V))$ proposed in this work, which are compatible with different results from the literature, are recommended for high-precision reddening correction in the future.

## 6. Summary

In this work, we have obtained accurate dust reddening from the far-UV to mid-IR for up to 5 million stars by the star-pair algorithm, using LAMOST DR7 and HotPayne stellar parameters along with GALEX, PS1, Gaia, SDSS, 2MASS, and WISE photometric data. The typical errors are approximately 0.01 mag or better for the PS1 and Gaia colors, and 0.02–0.03 mag for the SDSS and 2MASS and WISE colors (Figure 2). We have derived the empirical reddening coefficients for 21 colors both in the traditional (single-valued, Figure 4, Table 2) way and as a function of $T_{eff}$ and $E(B - V)$ (Figure 5, Tables 2 and 4), by using the largest sample stars of accurate reddening measurements together with the extinction values from the SFD all-sky reddening map. The corresponding extinction coefficients were also obtained (Table 3). The results were compared with model predictions (Figure 7) and the literature (Figure 8). The $T_{eff}$ and $E(B - V)$ dependent reddening coefficients $R(T_{eff}, E(B - V))$ provide a better description of empirical reddening coefficients than the single-valued ones. The broader the filters (e.g., the Gaia passbands) or the bluer the filters (e.g., the GALEX filters), the stronger the dependence. In most cases, the reddening coefficients are positively correlated with $E(B - V)$ and monotonically and inversely correlated with $T_{eff}$, which is consistent with the model predictions. Comparisons with measurements in the literature show that the $T_{eff}$- and $E(B - V)$-dependent coefficients can naturally explain the discrepancies between different measurements, i.e., using sample stars of different temperatures and reddening. Our coefficients are generally valid in the extinction range of 0–0.5 mag and the temperature range of 4000–10,000 K, and are recommended for high-precision reddening correction.

In the future, we will continue to use the multiband reddening values of millions of stars obtained in this work to study the 2D/3D variations of extinction laws in the Galaxy and to pave the way for high-precision reddening corrections.

We acknowledge the referee for the valuable comments that improved the clarity and quality of the manuscript. This work is supported by the National Key Basic R&D Program of China via 2019YFA0405500 and the National Natural Science Foundation of China through the projects NSFC 12222301 and 12173007. This work has made use of data products from the LAMOST, GALEX, PS1, Gaia, SDSS, 2MASS, and WISE. Guoshoujing Telescope (the Large Sky Area Multi-Object Fiber Spectroscopic Telescope, LAMOST) is a National Major Scientific Project built by the Chinese Academy of Sciences. Funding for the project has been provided by the National Development and Reform Commission. LAMOST is operated and managed by the National Astronomical Observatories, Chinese Academy of Sciences. We acknowledge the science research grants from the China Manned Space Project with Nos. CMS-CSST-2021-A08 and CMS-CSST-2021-A09.

## Appendix

In Figure 9 we show the distributions of 21 different intrinsic colors of the target samples in the stellar parameter space, which are the interim results of the DR7 catalog derived by the star-pair method. In general, intrinsic colors change smoothly with stellar parameters. The low-amplitude local wavy structures in the IR colors are due to the relatively large photometric errors of their control samples. The unnatural jagged boundaries at some edges are due to the elimination of target stars of too few control stars, as mentioned in Section 3.

Table 4 provides a quick check sheet of the reddening coefficients $R(T_{eff}, E(B - V))$ of each color at different $T_{eff}$ and $E(B - V)$ values.

We also provide a Python package to facilitate the querying of the extinction and reddening coefficients under different cases. Instructions on installation and usage of this package are available at https://github.com/vnohhf/extinction_coeffcient/.

**Table 4**
Reddening Coefficients at Different $T_{\rm eff}$ and $E(B-V)$

| Color | $T_{\rm eff}(K)$ | $E(B-V)$ (mag) | | | | | |
|---|---|---|---|---|---|---|---|
| | | 0 | 0.1 | 0.2 | 0.3 | 0.4 | 0.5 |
| FUV − NUV | 7500 | 0.498 | −0.099 | −0.527 | −0.784 | −0.871 | −0.787 |
| | 8500 | 0.912 | 0.315 | −0.113 | −0.370 | −0.457 | −0.373 |
| | 9500 | 1.379 | 0.782 | 0.354 | 0.097 | 0.010 | 0.094 |
| NUV − $g_{\rm PS}$ | 4500 | 0.462 | 0.161 | −0.147 | −0.463 | −0.785 | −1.114 |
| | 5250 | 2.836 | 2.535 | 2.227 | 1.911 | 1.589 | 1.259 |
| | 5750 | 3.923 | 3.622 | 3.313 | 2.998 | 2.676 | 2.346 |
| | 6500 | 4.979 | 4.678 | 4.369 | 4.054 | 3.732 | 3.402 |
| | 7500 | 5.670 | 5.369 | 5.060 | 4.745 | 4.423 | 4.093 |
| | 8500 | 6.037 | 5.736 | 5.427 | 5.112 | 4.790 | 4.460 |
| $g_{\rm PS}$ − $r_{\rm PS}$ | 4500 | 1.025 | 1.006 | 0.981 | 0.950 | 0.913 | 0.870 |
| | 5250 | 0.921 | 0.902 | 0.877 | 0.846 | 0.809 | 0.766 |
| | 5750 | 0.921 | 0.902 | 0.877 | 0.846 | 0.809 | 0.766 |
| | 6500 | 0.965 | 0.946 | 0.921 | 0.890 | 0.853 | 0.810 |
| | 7500 | 0.977 | 0.959 | 0.934 | 0.903 | 0.866 | 0.822 |
| $r_{\rm PS}$ − $i_{\rm PS}$ | 4500 | 0.705 | 0.677 | 0.653 | 0.631 | 0.612 | 0.596 |
| | 5250 | 0.626 | 0.599 | 0.574 | 0.552 | 0.533 | 0.517 |
| | 5750 | 0.621 | 0.594 | 0.570 | 0.548 | 0.529 | 0.513 |
| | 6500 | 0.646 | 0.619 | 0.594 | 0.573 | 0.554 | 0.537 |
| | 7500 | 0.651 | 0.623 | 0.599 | 0.577 | 0.558 | 0.542 |
| $i_{\rm PS}$ − $z_{\rm PS}$ | 4500 | 0.532 | 0.486 | 0.450 | 0.423 | 0.405 | 0.396 |
| | 5250 | 0.502 | 0.456 | 0.420 | 0.393 | 0.375 | 0.366 |
| | 5750 | 0.496 | 0.450 | 0.414 | 0.387 | 0.369 | 0.360 |
| | 6500 | 0.501 | 0.455 | 0.419 | 0.392 | 0.374 | 0.365 |
| | 7500 | 0.517 | 0.472 | 0.435 | 0.408 | 0.390 | 0.381 |
| | 8500 | 0.522 | 0.476 | 0.440 | 0.413 | 0.395 | 0.386 |
| $z_{\rm PS}$ − $y_{\rm PS}$ | 4500 | 0.315 | 0.301 | 0.288 | 0.277 | 0.267 | 0.259 |
| | 5250 | 0.287 | 0.273 | 0.260 | 0.248 | 0.239 | 0.230 |
| | 5750 | 0.282 | 0.268 | 0.255 | 0.244 | 0.234 | 0.225 |
| | 6500 | 0.288 | 0.274 | 0.261 | 0.250 | 0.240 | 0.232 |
| | 7500 | 0.304 | 0.290 | 0.277 | 0.265 | 0.256 | 0.247 |
| | 8500 | 0.304 | 0.290 | 0.277 | 0.266 | 0.256 | 0.247 |
| $y_{\rm PS}$ − $J$ | 4500 | 0.426 | 0.464 | 0.485 | 0.489 | 0.475 | 0.445 |
| | 5250 | 0.335 | 0.374 | 0.395 | 0.398 | 0.385 | 0.354 |
| | 5750 | 0.321 | 0.359 | 0.380 | 0.384 | 0.370 | 0.340 |
| | 6500 | 0.329 | 0.367 | 0.388 | 0.392 | 0.379 | 0.348 |
| | 7500 | 0.313 | 0.351 | 0.372 | 0.376 | 0.362 | 0.331 |
| $J − H$ | 4500 | 0.402 | 0.372 | 0.348 | 0.331 | 0.321 | 0.316 |
| | 5250 | 0.370 | 0.340 | 0.316 | 0.299 | 0.289 | 0.284 |
| | 5750 | 0.362 | 0.332 | 0.308 | 0.291 | 0.280 | 0.276 |
| | 6500 | 0.358 | 0.328 | 0.305 | 0.287 | 0.277 | 0.273 |
| | 7500 | 0.346 | 0.316 | 0.292 | 0.275 | 0.264 | 0.260 |
| $H − Ks$ | 4500 | 0.101 | 0.119 | 0.131 | 0.135 | 0.133 | 0.123 |
| | 5250 | 0.113 | 0.131 | 0.143 | 0.147 | 0.144 | 0.135 |
| | 5750 | 0.117 | 0.136 | 0.147 | 0.151 | 0.149 | 0.139 |
| | 6500 | 0.120 | 0.138 | 0.149 | 0.154 | 0.151 | 0.142 |
| | 7500 | 0.119 | 0.137 | 0.148 | 0.153 | 0.150 | 0.141 |
| $Ks − W1$ | 4500 | 0.121 | 0.122 | 0.121 | 0.117 | 0.112 | 0.104 |
| | 5250 | 0.118 | 0.119 | 0.118 | 0.114 | 0.109 | 0.101 |
| | 5750 | 0.116 | 0.117 | 0.116 | 0.112 | 0.107 | 0.099 |
| | 6500 | 0.114 | 0.115 | 0.114 | 0.110 | 0.104 | 0.096 |
| | 7500 | 0.115 | 0.115 | 0.114 | 0.111 | 0.105 | 0.097 |
| W1-W2 | 4500 | 0.097 | 0.086 | 0.078 | 0.073 | 0.070 | 0.069 |
| | 5250 | 0.082 | 0.072 | 0.064 | 0.059 | 0.056 | 0.055 |
| | 5750 | 0.082 | 0.072 | 0.064 | 0.058 | 0.055 | 0.054 |
| | 6500 | 0.083 | 0.073 | 0.065 | 0.059 | 0.056 | 0.055 |
| | 7500 | 0.063 | 0.052 | 0.045 | 0.039 | 0.036 | 0.035 |

**Table 4**
(Continued)

| Color | $T_{eff}(K)$ | E(B − V) (mag) | | | | | |
|---|---|---|---|---|---|---|---|
| | | 0 | 0.1 | 0.2 | 0.3 | 0.4 | 0.5 |
| BP − G | 4500 | 0.717 | 0.707 | 0.695 | 0.682 | 0.668 | 0.653 |
| | 5250 | 0.658 | 0.647 | 0.635 | 0.623 | 0.609 | 0.594 |
| | 5750 | 0.629 | 0.619 | 0.607 | 0.594 | 0.580 | 0.566 |
| | 6500 | 0.600 | 0.589 | 0.577 | 0.564 | 0.551 | 0.536 |
| | 7500 | 0.576 | 0.565 | 0.553 | 0.541 | 0.527 | 0.512 |
| | 8500 | 0.559 | 0.548 | 0.536 | 0.523 | 0.510 | 0.495 |
| | 9500 | 0.536 | 0.525 | 0.514 | 0.501 | 0.487 | 0.472 |
| G − RP | 4500 | 0.585 | 0.555 | 0.524 | 0.490 | 0.454 | 0.417 |
| | 5250 | 0.655 | 0.625 | 0.594 | 0.560 | 0.524 | 0.487 |
| | 5750 | 0.704 | 0.675 | 0.643 | 0.610 | 0.574 | 0.536 |
| | 6500 | 0.779 | 0.749 | 0.718 | 0.684 | 0.648 | 0.611 |
| | 7500 | 0.871 | 0.842 | 0.810 | 0.776 | 0.741 | 0.703 |
| | 8500 | 0.944 | 0.915 | 0.883 | 0.849 | 0.814 | 0.776 |
| | 9500 | 0.986 | 0.957 | 0.925 | 0.891 | 0.856 | 0.818 |
| RP − Ks | 4500 | 1.541 | 1.534 | 1.513 | 1.478 | 1.428 | 1.365 |
| | 5250 | 1.472 | 1.465 | 1.444 | 1.409 | 1.360 | 1.297 |
| | 5750 | 1.456 | 1.448 | 1.427 | 1.392 | 1.343 | 1.280 |
| | 6500 | 1.461 | 1.454 | 1.433 | 1.397 | 1.348 | 1.285 |
| | 7500 | 1.499 | 1.492 | 1.470 | 1.435 | 1.386 | 1.323 |
| | 8500 | 1.535 | 1.528 | 1.506 | 1.471 | 1.422 | 1.359 |
| $u_{sdss} − g_{sdss}$ | 4500 | 1.132 | 1.018 | 0.925 | 0.854 | 0.804 | 0.776 |
| | 5250 | 1.200 | 1.085 | 0.992 | 0.921 | 0.871 | 0.844 |
| | 5750 | 1.239 | 1.125 | 1.032 | 0.961 | 0.911 | 0.883 |
| | 6500 | 1.288 | 1.174 | 1.081 | 1.010 | 0.960 | 0.932 |
| | 7500 | 1.329 | 1.214 | 1.122 | 1.050 | 1.001 | 0.973 |
| | 8500 | 1.334 | 1.220 | 1.127 | 1.056 | 1.006 | 0.978 |
| $g_{sdss} − r_{sdss}$ | 4500 | 1.250 | 1.203 | 1.156 | 1.109 | 1.063 | 1.017 |
| | 5250 | 1.126 | 1.079 | 1.033 | 0.986 | 0.940 | 0.894 |
| | 5750 | 1.124 | 1.077 | 1.030 | 0.984 | 0.938 | 0.891 |
| | 6500 | 1.168 | 1.121 | 1.075 | 1.028 | 0.982 | 0.936 |
| | 7500 | 1.165 | 1.118 | 1.071 | 1.025 | 0.978 | 0.932 |
| $r_{sdss} − i_{sdss}$ | 4500 | 0.758 | 0.709 | 0.671 | 0.644 | 0.626 | 0.618 |
| | 5250 | 0.685 | 0.637 | 0.599 | 0.571 | 0.553 | 0.546 |
| | 5750 | 0.672 | 0.624 | 0.586 | 0.558 | 0.540 | 0.533 |
| | 6500 | 0.681 | 0.633 | 0.595 | 0.567 | 0.549 | 0.542 |
| | 7500 | 0.692 | 0.644 | 0.606 | 0.578 | 0.560 | 0.552 |
| $i_{sdss} − z_{sdss}$ | 4500 | 0.730 | 0.634 | 0.564 | 0.521 | 0.504 | 0.513 |
| | 5250 | 0.690 | 0.594 | 0.524 | 0.481 | 0.464 | 0.473 |
| | 5750 | 0.680 | 0.583 | 0.514 | 0.470 | 0.453 | 0.462 |
| | 6500 | 0.678 | 0.582 | 0.512 | 0.469 | 0.451 | 0.460 |
| | 7500 | 0.686 | 0.590 | 0.520 | 0.476 | 0.459 | 0.468 |
| | 8500 | 0.680 | 0.583 | 0.513 | 0.470 | 0.453 | 0.462 |
| $z_{sdss} − Ks$ | 4500 | 1.010 | 1.083 | 1.122 | 1.127 | 1.099 | 1.036 |
| | 5250 | 0.878 | 0.951 | 0.990 | 0.996 | 0.967 | 0.905 |
| | 5750 | 0.864 | 0.937 | 0.976 | 0.982 | 0.953 | 0.891 |
| | 6500 | 0.903 | 0.976 | 1.016 | 1.021 | 0.992 | 0.930 |
| | 7500 | 0.963 | 1.036 | 1.075 | 1.080 | 1.052 | 0.989 |

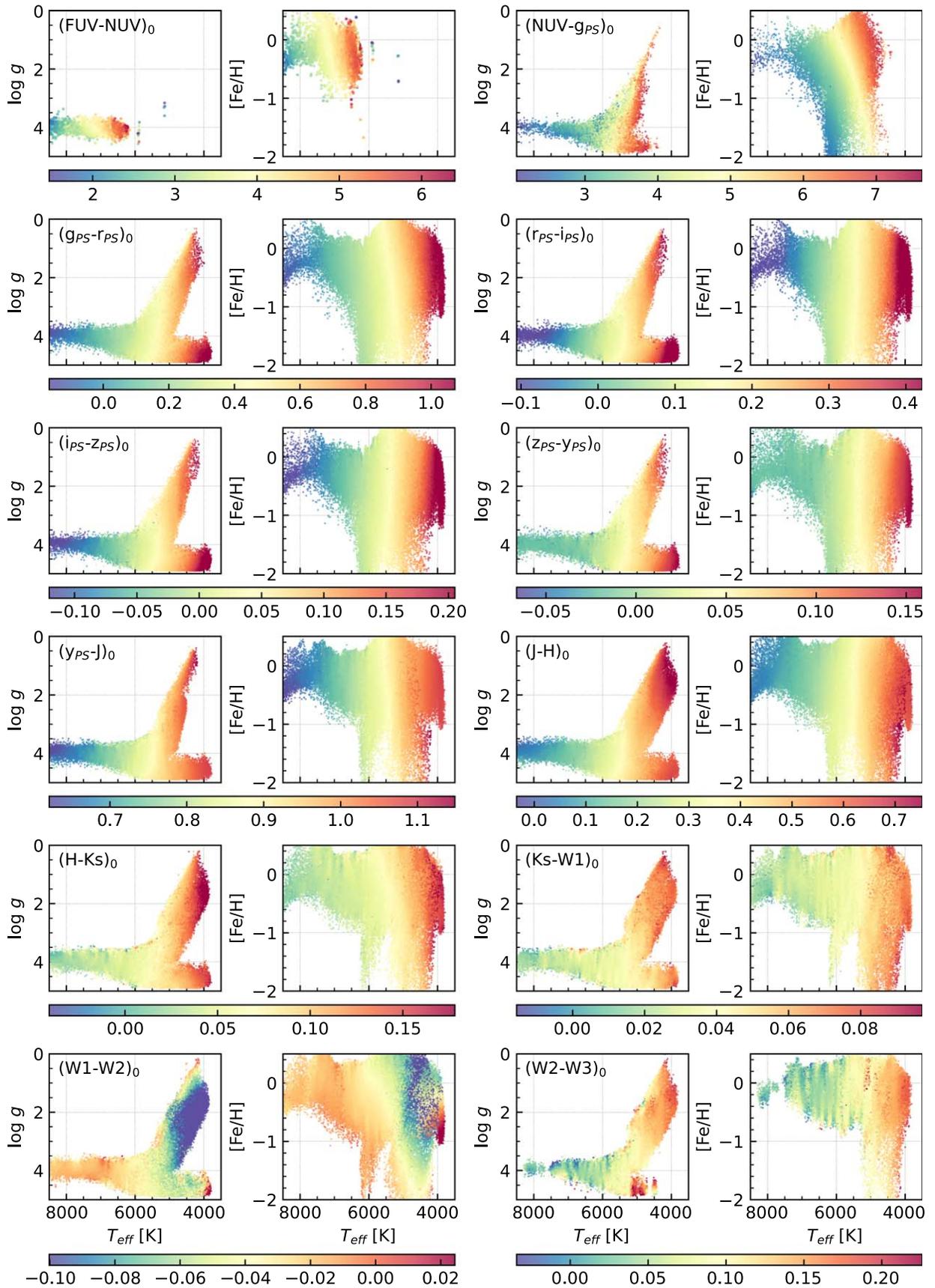

**Figure 9.** Distributions of different intrinsic colors derived by the star-pair method in the stellar parameter space. The color-coded dots represent the target samples in the LAMOST DR7 catalog. The color bars represent the stellar colors of the filter combination labeled on the plots. Note that only a randomly selected 10% of the target sample is drawn.

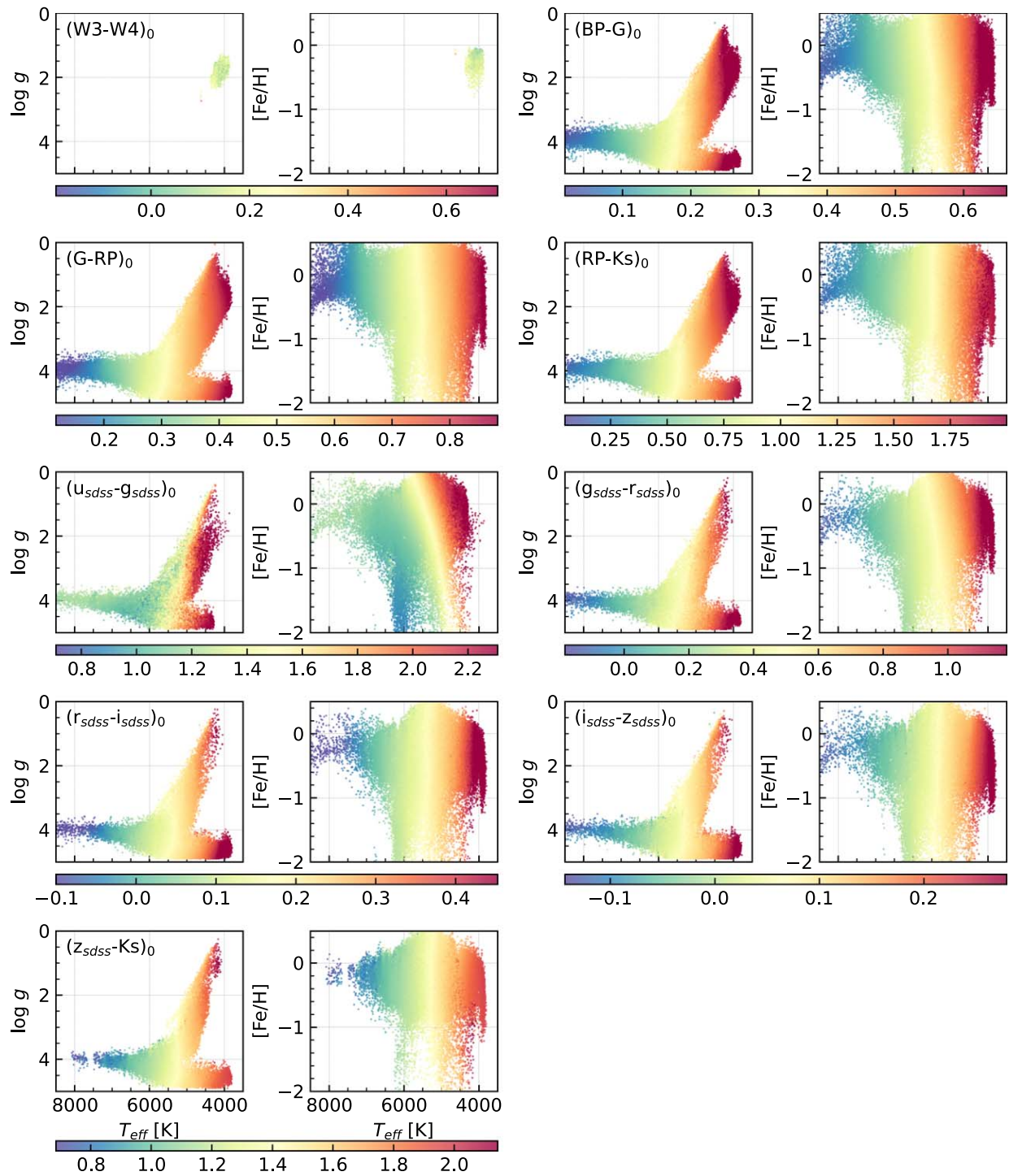

**Figure 9.** (Continued.)

**ORCID iDs**

Ruoyi Zhang (iD) https://orcid.org/0000-0003-1863-1268
Haibo Yuan (iD) https://orcid.org/0000-0003-2471-2363